\documentclass[preprint,floats,aps,epsfig,nofootinbib,amssymb]{revtex4}

\def\lsim{\mathrel{\rlap{\lower4pt\hbox{\hskip1pt$\sim$}}
    \raise1pt\hbox{$<$}}}         
\def\gsim{\mathrel{\rlap{\lower4pt\hbox{\hskip1pt$\sim$}}
    \raise1pt\hbox{$>$}}}         
\def\R{\mathbb {S}}

\usepackage{slashed}
\usepackage{slashed}
\usepackage{graphicx,color}
\usepackage{epsfig}
\usepackage{subfigure}
\usepackage{epsfig}
\usepackage{dcolumn}
\usepackage{bm}
\usepackage{color}

\begin{document}

\vspace*{-5.8ex}
\hspace*{\fill}{ACFI-T15-20}

\vspace*{+3.8ex}

\title{Neutrino Catalyzed  Diphoton Excess}

\author{Wei Chao}
\email{chao@physics.umass.edu}

 \affiliation{ Amherst Center for Fundamental Interactions, Department of Physics, University of Massachusetts-Amherst
Amherst, MA 01003 }

\vspace{3cm}

\begin{abstract}

In this paper we explain the 750 GeV diphoton resonance observed at the run-2 LHC  as a scalar singlet $\R$, that plays a key rule in generating tiny but nonzero Majorana neutrino masses. 
The model contains four electroweak singlets: two leptoquarks, a singly charged scalar and a neutral scalar $\R$. 
Majorana neutrino masses might be generated at the two-loop level as $\R$ get nonzero vacuum expectation value. 
$\R$ can be produced at the LHC through the gluon fusion and decays into diphoton at the one-loop level with charged scalars running in the loop. 
The model fits perfectly with a wide width of the resonance. 
Constraints on the model are investigated, which shows a negligible mixing between the resonance and the standard model Higgs boson.

\end{abstract}

\maketitle
\section{Introduction}

The standard model (SM) of particle physics fits perfectly with almost all the experimental observations in the elementary particle physics.
But there are still hints of new physics beyond the SM. 
The discovery of the neutrino oscillations has confirmed that neutrinos are massive and lepton flavors are mixed~\cite{Agashe:2014kda}, which provided the first evidence for physics beyond the SM. 
An attractive approach towards understanding the origin of small neutrino masses is using the dimension-five Weinberg operator~\cite{Weinberg:1979sa}
\begin{eqnarray}
{1\over 4 } \kappa_{gf}^{} \overline{\ell_{Lc}^C}^g \varepsilon_{cd} H_d^{} \ell_{Lb}^f \varepsilon_{ba} H_a^{}  + {\rm h.c.} 
\end{eqnarray}
where $\ell_L$ is the left-handed lepton doublet, $H$ is the SM Higgs. 
The operator, which might come from integrating out some new superheavy particles,  gives Majorana masses to active neutrinos after the spontaneous breaking of the electroweak symmetry.  
A simple way to realize the Weinberg operator is through the tree-level seesaw mechanisms~\cite{seesawI,seesawII,seesawIII}. 
But the canonical seesaw scales are usually too high to be accessible by colliders. 
Many TeV-scale seesaw mechanisms were proposed motivated by the testability, some of which give rise to the Weinberg operator at the loop level.

Recently both the ATLAS~\cite{ATLAS13} and CMS~\cite{CMS13} collaborations have observed a new resonance at an invariant mass of $750$ GeV in the diphoton channel at the run-2 LHC with center-of-mass energy $\sqrt{s}=13~{\rm TeV}$.  
The local significance is $3.6\sigma$ and $2.6\sigma$ for ATLAS and CMS respectively. 
The best fit width of the resonance from the ATLAS is about $45~{\rm GeV}$ and the corresponding cross section is about $10\sim 3$ fb, while the CMS result favors the narrow width, which has $\sigma(pp\to \gamma \gamma ) \approx 2.6\sim 7.7~{\rm fb}$ with $\Gamma\approx0.1~{\rm GeV}$.
If confirmed, it would be another hint of new physics beyond the SM.
According to Landau-Yang theorem~\cite{Landau:1948kw,Yang:1950rg}, this resonance can only be spin-0 or spin-2 bosonic state. 
The heavy quarks and/or gluon fusion production of the resonance are favored, because the run-1 LHC~\cite{ATLAS8,CMS8} at $\sqrt{s}=8~{\rm TeV}$ did not see significant excess at the 750 GeV.
It is intriguing to investigate the physics behind this excess as done in a bunch of papers~~\cite{Angelescu:2015uiz,Backovic:2015fnp,Bellazzini:2015nxw,Buttazzo:2015txu,DiChiara:2015vdm,Ellis:2015oso,Franceschini:2015kwy,Gupta:2015zzs,Harigaya:2015ezk,Higaki:2015jag,Knapen:2015dap,Low:2015qep,Mambrini:2015wyu,McDermott:2015sck,Molinaro:2015cwg,Nakai:2015ptz,Petersson:2015mkr,Pilaftsis:2015ycr,Dutta:2015wqh,Cao:2015pto,Matsuzaki:2015che,Kobakhidze:2015ldh,Martinez:2015kmn,Cox:2015ckc,Becirevic:2015fmu,No:2015bsn,Demidov:2015zqn,Chao:2015ttq,Fichet:2015vvy,Curtin:2015jcv,Bian:2015kjt,Chakrabortty:2015hff,Agrawal:2015dbf,Csaki:2015vek,Falkowski:2015swt,Aloni:2015mxa,Bai:2015nbs,Chao:2015nsm,Han:2015cty}. 
Especially it would be more interesting if a new physics can explain both the diphoton excess and other unsolved problems of the SM, such as neutrino masses and dark matter.

In this paper,  we explain both the diphoton excess and the active neutrino masses in a concrete model.
The model extends the SM with four scalar singlet: one neutral scalar $\R$, one singly charged scalar $\Theta$ and two leptoquarks $\Phi$ and $\Omega$ transforming as $(\bar3,~1, ~4/3)$ and $(\bar3,~1,~1/3)$ under the SM gauge group $SU(3)_C^{}  \times SU(2)_L^{} \times U(1)_Y^{} $.  
$\Theta$ interacts with the left-handed lepton doublets,  $\Omega$ interacts with the left-handed lepton- and quark- doublets, $\Phi$ interacts with the  right-handed leptons and down-type quarks. 
In addition there is a quartic interaction of the type $  \hat \lambda \R \Phi^\dagger \Omega \Theta +{\rm h.c.} $.  
In this way  Majorana neutrino masses can be generated at the two-loop level as $\R$ gets vacuum expectation value (VEV).  
Elements of neutrino mass matrix are proportional to the charged lepton and down-type quark masses, and are suppressed by the loop factor. 
Thus one can naturally derive the electro-volt scale Majorana neutrino masses with new particles at the TeV scale. 
Furthermore, the observed 750 GeV diphoton excess can be explained as the scalar $\R$, which is produced through the gluon fusion at the LHC.   
The ratio of $\Gamma(\R \to gg)/\Gamma(\R \to \gamma \gamma)$ is of ${\cal O }(25)$, which is the  typical character of this model. 
The wide width of $\R$ can be explained if either charged scalars are sightly lighter than $m_\R/2$ or $\R$ is the mediator of the hidden valley.  
We further point out that $\R$ might play important rule in the spontaneous breaking of a local $U(1)_{B+L}$ gauge symmetry, which was already studied in Ref.~\cite{Chao:2015nsm}.  
Constraints on the model from the run-1 LHC are studied. 
Leptoquarks are supposed to mainly couple to the third generation, or couple to the hidden valley, so as to escape the collider search constraints. 
The mixing angle of the resonance with the SM Higgs is constrained as $\theta <0.067$, assuming $\Gamma(\R\to \gamma\gamma) \sim 0.05~{\rm GeV}$.  
Furthermore the model may have gauge couplings unification by introducing an extra quadruplet scalar.

The remaining of the paper is organized as follows: In section II we study how to generate Majorana neutrino masses at the two-loop level. 
In section III, we investigate the possibility of explaining the diphoton excess as the scalar $\R$ in our model.  
Constraints and further predictions are given in section IV.
The last part is concluding remarks.

\section{Neutrino mass}

In the SM neutrinos are massless, both the lepton number and lepton flavors are conserved.  
Reactor, solar, atmosphere and accelerator neutrino experiments have discovered neutrino oscillations, which means neutrinos have tiny but nonzero masses.  
The origin of neutrino masses is still an open question.
Seesaw mechanisms are probably the most natural way to understand the smallness of neutrino masses. 
But elegant traditional seesaw mechanisms are not accessible by colliders, because either the seesaw scales are too high or the couplings of the seesaw particles with the SM fields are too tiny. 
Motivated by the testability, many TeV-scale seesaw models~\cite{Wei:2010ww,Angel:2013hla,Sierra:2014rxa} were proposed.
In this section we propose a new TeV-scale seesaw model, which generate neutrino masses at the two-loop level.
The new seesaw model extends the SM with four electroweak singlets: two scalar leptoquarks $\Phi$ and $\Omega$ transforming as $(\bar{3},~1,~{4/3})$ and $(\bar{3},~1, ~1/3)$ respectively under the SM gauge group $SU(3)_C\times SU(2)_L \times U(1)_Y$, one singly charged scalar $\Theta$ transforming as $( 1,~1,~1)$ and a neutral scalar $\R$ transforming as $(1,~1,~0)$.  
New interactions can be written as
\begin{eqnarray}
{\cal L} \supset  \hat{Y}_\Theta^{} {\ell_L^T} i\sigma_2^{} \ell_L^{} \Theta  + \hat{Y}_\Omega^{} {q_L^T} i\sigma \ell_L^{} \Omega+ \hat{Y}_\Phi^{} E_R^{T} d_R^{} \Phi -V(\R,\Phi, \Omega, \Theta) \label{yukawa}
\end{eqnarray}
where $\ell_L$ is the left-handed lepton doublet, $q_L $ is the left-handed quark doublet, $E_R^{}$ and $d_R^{}$ are right-handed charged leptons and down-type quarks respectively.   
The new scalar potential takes the form
\begin{eqnarray}
V(\R,\Phi, \Omega, \Theta ) &\supset& -\mu_\R^2 \R^\dagger \R + \lambda_\R (\R^\dagger \R)^2 + \mu_\Theta^2 \Theta^\dagger \Theta + \mu_\Phi^2 \Phi^\dagger \Phi + \mu_\Omega^2 \Omega^\dagger \Omega + g_{\Theta\R}^{}  (\Theta^\dagger \Theta ) (\R^\dagger \R)  \nonumber \\
&& +g_{\Phi\R}^{}  (\Phi^\dagger \Phi ) (\R^\dagger \R) +g_{\Omega\R}^{}  (\Omega^\dagger \Omega ) (\R^\dagger \R) +\{ {\sqrt{2}\hat \lambda \R \Theta  \Phi^\dagger \Omega + {\rm h.c.}}\}+ \cdots \label{potential}
\end{eqnarray}
where  dots denotes terms that are irrelevant to our study.  Notice that interactions in Eqs. (\ref{yukawa}) and (\ref{potential})  have a global U(1) lepton number  (L) symmetry by setting $L_\R=-2$. 
It  might be spontaneously broken as $\R$ gets non-zero vacuum expectation value (VEV), $\mathbf{v}_\R$, at the TeV scale. 
For the case $\R$ being a complex scalar, there will be a massless goldstone boson, which is severely constrained by the big bang nucleosynthesis~\cite{Barger:2003zg} and observations of Bullet Cluster galaxies~\cite{Randall:2007ph}.
To avoid this problem, one might add the soft $U(1)_L$  breaking mass term i.e. $\mu_R^{\prime 2}(\R^2 +{\rm h.c.}) $ to the potential without introducing any other trouble. 
Alternatively,  if $\R$ itself is a real scalar singlet, there will be no massless goldstone boson.
Actually, this model can be embedded into a theory with local $U(1)_{B+L}$ gauge symmetry\footnote{We only point out this possibility, but not focus on this scenario in the rest of the paper.}, where anomalies might be cancelled by introducing colorless vector-like fermions~\cite{Chao:2010mp}.  
For the systematic studies of local  $U(1)_B\times U(1)_L$ gauge symmetries, we refer the reader to Refs. \cite{FileviezPerez:2010gw,FileviezPerez:2011pt,Duerr:2013dza} for detail.
After $\R$ getting non-zero VEV, the mass eigenvalues of scalars can be written as
\begin{eqnarray}
M_\Theta^2 = \mu_\Theta^2 + {1\over 2 }g_{\Theta \R }^{} \mathbf{v}_\R^2 \; ,  \hspace{0.5cm} M_\Omega^2 = \mu_\Omega^2 + {1\over 2 } g_{\Omega \R}^{} \mathbf{v}_\R^2 \; \hspace{0.5cm} M_\Phi^2 = \mu_\Phi^2 + {1\over 2 } g_{\Phi \R}^{} \mathbf{v}_\R^2 \; .
\end{eqnarray}
%

\begin{figure}
  \includegraphics[width=0.45\textwidth]{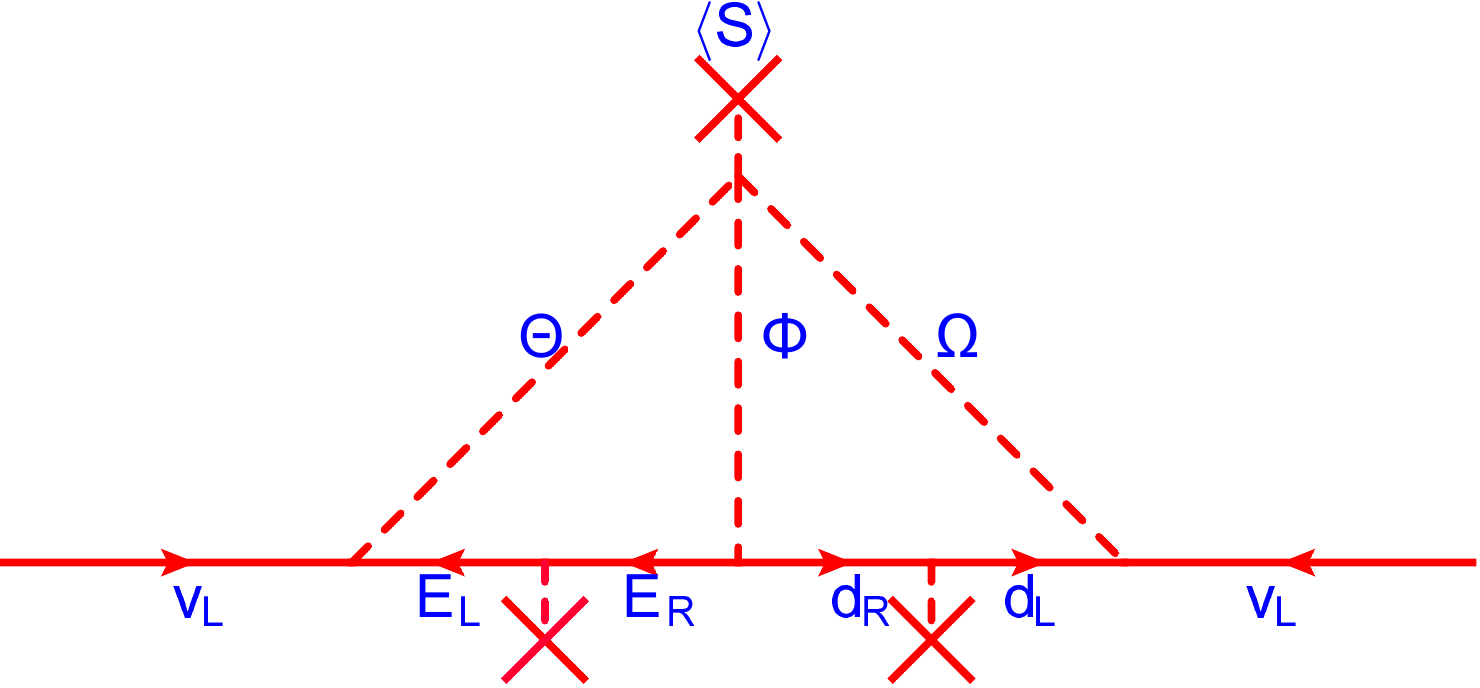}
\caption{\label{feyn}  Feynman diagram that generates neutrino masses at the two-loop level.}
\end{figure}

Given the Yukawa interactions in Eq. (\ref{yukawa}), neutrino masses can be generated at the two-loop level. 
The relevant feynman diagram is given in Fig. \ref{feyn}.  Neutrino mass elements can be written in terms of loop functions and Yukawa couplings 
\begin{eqnarray}
( M_\nu^{} )_{ij}^{} &=& {1\over 1+\delta_{ij}}\sum_{kl} \left[ (\hat{Y}_{\Theta}^{} )_{ik}  ( \hat{Y}_{\Phi}^{*} )_{kl}^{} (\hat{Y}_\Omega)_{lj}^{}  + (\hat{Y}_{\Theta}^{} )_{j k}    (\hat{Y}_{\Phi}^{*} )_{kl}^{} (\hat{Y}_\Omega)_{li}^{}  \right ] \hat{\lambda } \mathbf{v}_\R  \hat m^d_{l}  \hat m^e_k I_{kl}^{}   \label{neutrinomass}
\end{eqnarray}
where $m^e_k$ and $m^d_l$ are mass eigenvalues of the charged leptons and down-type quarks respectively, 
$I_{kl}$ is the loop function, which can be written as
\begin{eqnarray}
I_{kl}^{} = \int {d^4 k \over (2\pi)^4 } \int {d^4 l\over (2\pi)^4} {1\over P_\Theta^2 -M_\Theta^2 }  {1\over P_\Omega^2 -M_\Omega^2 }  {1\over P_\Phi^2 -M_\Phi^2 }  {1\over P_{k}^2 -(m^{e}_k)^2 }  {1\over P_{d}^2 -(m_l^{d})^2 } 
\end{eqnarray}
We refer the reader to Refs.~\cite{Angel:2013hla,Sierra:2014rxa} for the calculation of this integral in detail.  
Apparently neutrino masses, generated through Eq. (\ref{neutrinomass}), are suppressed by the charged lepton and down-type quark masses as well as the loop factor.
All seesaw particles can be at the electroweak scale and the model is detectable at the LHC.  
A  systematic study the neutrino phenomenology and collider signatures  induced by this model, which are interesting but beyond the reach of this paper, will be given in a longer paper. 
Here we only estimate the size of parameters in Eq. (\ref{neutrinomass}) constrained by the active neutrino masses. 
The integral $I_{kl}^{}$ was calculated analytically in Ref.~\cite{McDonald:2003zj}, which has 
\begin{eqnarray}
I_{kl}^{} \propto {1\over 256 \pi^4} {1\over M^2 } {\pi^2 \over 3 }  \tilde I, \hspace{0.5cm}  M\equiv {\rm max} ( M_\Theta, M_\Omega, M_\Phi )
\end{eqnarray}
where $\tilde I \sim 1 $ for a large range of scalar masses.
Assuming $M\sim 400~{\rm GeV}$, $v_\R\sim 1~ {\rm TeV}$ and $\hat\lambda\sim 0.1$, one has $\hat Y_X\sim {\cal O} (0.1) (X=\Theta, \Omega, \Phi)$ to generate active neutrino masses at the sub-eV scale .

\begin{figure}
  \includegraphics[width=0.45\textwidth]{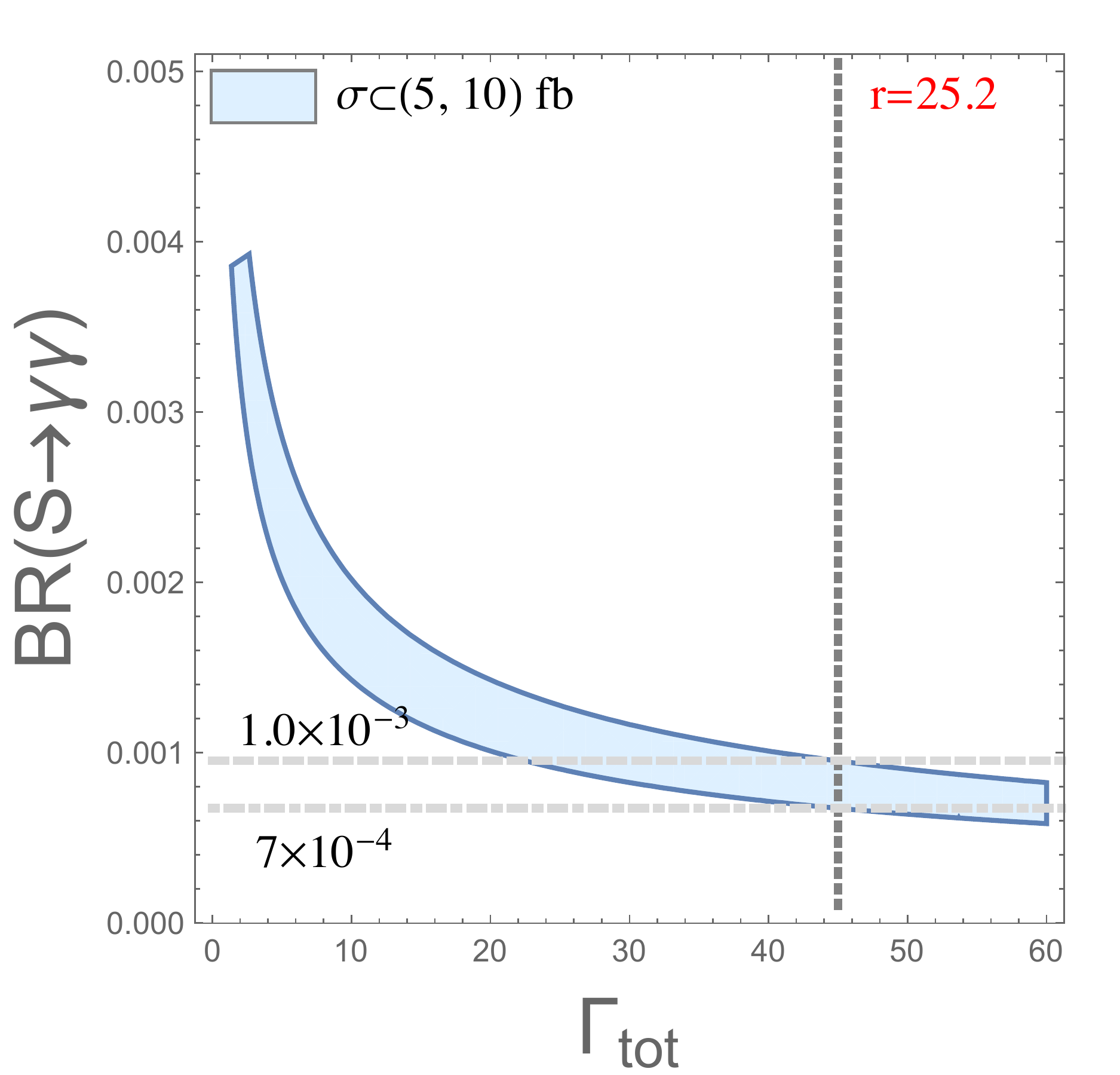}
\caption{\label{region}  The region in the $\Gamma_{\rm tot}^\R-{\rm BR} (\R \to \gamma \gamma )$ plane that has $\sigma (gg\to \R \to \gamma \gamma ) \in (5, ~10)$ fb, the vertical line has $\Gamma_{tot}^{\R} =45~{\rm GeV}$; the dot-dashed and dashed horizontal lines correspond to ${\rm Br} (\R \to \gamma \gamma ) =7\times 10^{-4}$ and $1\times 10^{-3}$ respectively.}
\end{figure}

\section{LHC diphoton excess}

Both the CMS and ATLAS experiments have observed the diphoton excess at $m_{\gamma\gamma} =750~{\rm GeV}$.  
The cross section is estimated as $\sigma(pp\to \gamma \gamma ) \approx  (6\pm3)~{\rm fb}$ for CMS~\cite{CMS13} and $(10\pm 3) ~{\rm fb }$ for ATLAS~\cite{ATLAS13} at the  $\sqrt{s}=13~{\rm TeV}$. 
Besides, the best fit width is about $45~{\rm GeV}$ from the ATLAS result, while the CMS result favors  the narrow width scenario.
If confirmed, the excess would be a solid evidence of new physics beyond the SM. 
In this section, we explain this resonance as  the scalar $\R$, that plays the key rule in generating active neutrino masses at the two-loop level.    
Furthermore,  $\R$ might also be the particle breaking the local $U(1)_{B+L}$ gauge symmetry spontaneously though the Higgs mechanism~\cite{Chao:2015nsm}. 
In this way the diphoton resonance might be a hint of new symmetries beyond the SM. 
The signal of  the  resonance at the LHC is 
\begin{eqnarray}
\sigma (pp\to {\R}\to \gamma \gamma ) = {1 \over  s}  { \Gamma_{tot}^{\R} \over M} C_{gg}^{} {\rm BR } (\R\to gg ) {\rm BR} (\R\to \gamma \gamma) \; , \label{basic}
\end{eqnarray}
 where $\sqrt{s}$ is the centre-of-mass energy,  $M$ is the mass of $\R$,  $ C_{gg}$ and $C_{q\bar q}$ are dimensionless partonic integrals, ${\rm BR} (\R\to XX)$  and $\Gamma_{\rm tot}^\R$ are the branching ratio and total decay rate of $\R$. 
Considering $C_{gg} \gg C_{q \bar q}$ and ${\rm BR}(\R \to q \bar q )$ is loop suppressed in our model, gluon fusion turns out to be the dominant production channel at the LHC.

The decay rate of $\R$ can be written as
\begin{eqnarray}
\Gamma (\R \to g g ) &=& {\alpha_s^2 m_\R^3 \over 128 \pi^3 } \left| \sum_X {1\over 2 }{g_X  \mathbf{v}_\R\over  M_X^2  }  A^{} (\tau_S)  \right|^2  \label{stogg}\\
\Gamma (\R \to \gamma \gamma ) &=& {\alpha^2 m_\R^3 \over 1024 \pi^3 } \left| \sum_X n^c_X Q_X^2 {g_X  \mathbf{v}_\R\over  M_X^2  }  A^{} (\tau_S)  \right|^2 
\end{eqnarray}
where the factor $1/2$ in the Eq. (\ref{stogg}) comes from ${\rm Tr} [\lambda^a \lambda^b] $, with $\lambda^a $ $(a=1\cdots8)$ the generators of  $SU(3)_c$; $\tau_X \equiv 4 M_X^2 /M_\R^2$.
%
The loop function can be written as~\cite{Carena:2012xa}
\begin{eqnarray}
A(x) &=& x-x^2 f(x) 
\end{eqnarray}
where $f(x)\equiv \arcsin^2(\sqrt{1/x} )$ by assuming $2M_X > M_\R$. We refer the reader to Ref.~\cite{Carena:2012xa} for the expression of $f(x)$ when $2M_X <M_\R$. 
There are other decay channels of $\R$, such as  invisible decay, which is relevant to fitting the total width and will  be discussed later, and four fermions cascade decays, which are suppressed by the phase space.

Two interesting scenarios will be discussed: scenario (i) $\R$ is a real singlet and there is no extra gauge symmetry;  scenario (ii) $\R$ is  a complex singlet that triggers the spontaneous breaking of the local $U(1)_{B+L}$ gauge symmetry. 
We mainly focus on the scenario (i) and will briefly comment on the scenario (ii).
We further assume $\Phi$, $\Omega$ and $\Theta$ have degenerate masses, $M_X$, and the same couplings, $g_X$, with the $\R$ just for simplification.   
%

\begin{figure}
  \includegraphics[width=0.45\textwidth]{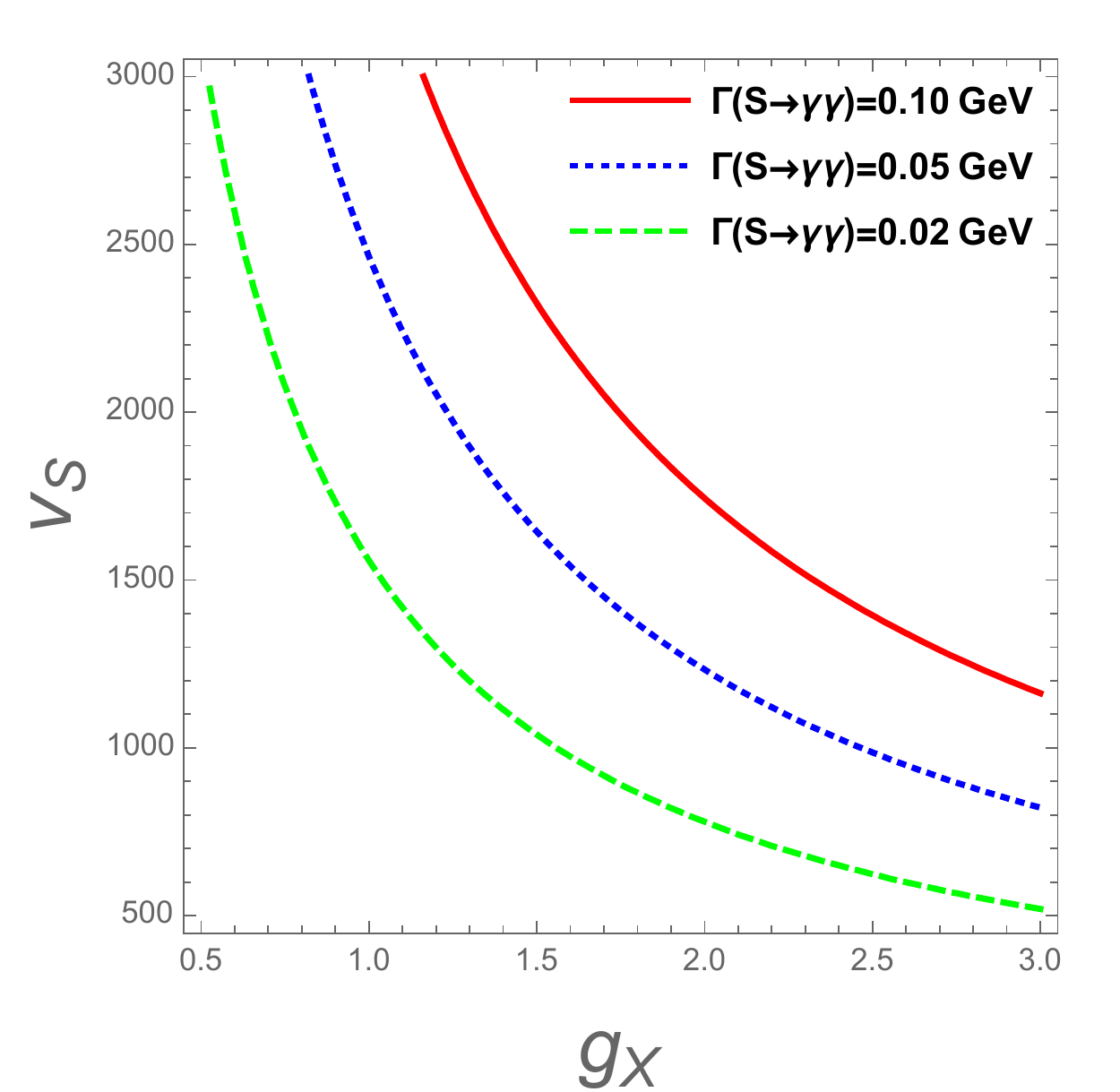}
   \includegraphics[width=0.45\textwidth]{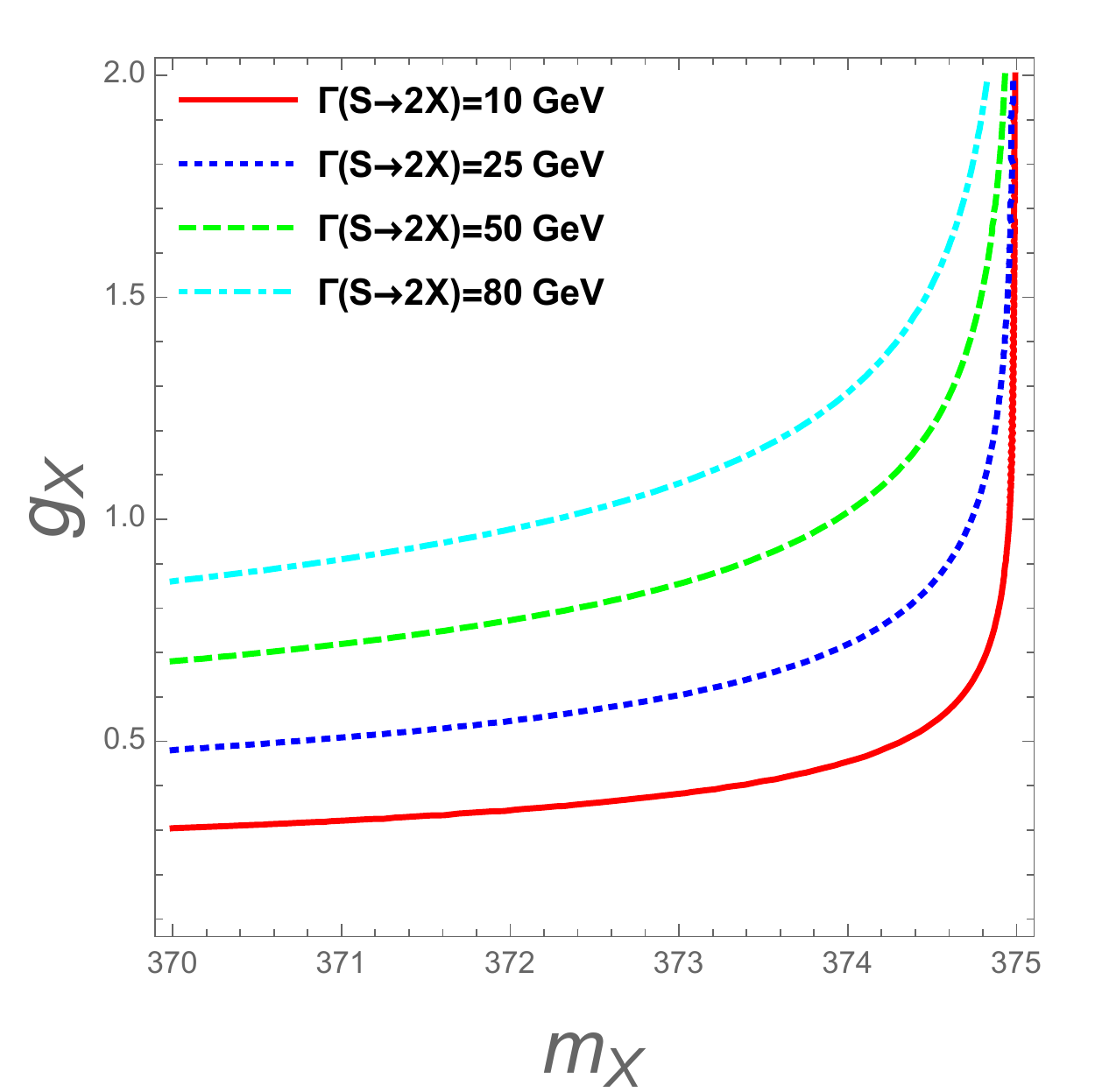}
\caption{\label{widthi}  Left panel: contours of the $\Gamma(\R \to \gamma \gamma ) $ in the $g_X$-$v_S$ plane, where $g_X$ is the universal coupling between $\R$ and other new scalars; Right panel: contours of $\Gamma (\R \to 2 X)$ in the $M_X -g_X $ plane by setting $\mathbf{v}_\R=2.5~{\rm TeV}$.}
\end{figure}

For scenario (i), one has
\begin{eqnarray}
r \equiv {\Gamma (\R \to gg ) \over \Gamma (\R \to \gamma \gamma )} = {2 \alpha_s^2  \over  \alpha^2} \left({3\over 20 } \right)^2 \approx  25.2
\end{eqnarray}
by setting $\alpha_s \approx0.0934$ and $\alpha \approx 1/126.8$. It is the typical character of this scenario. 
Given this typical input, we plot in the $\Gamma^{\R}_{\rm tot} - {\rm Br} (\R \to \gamma \gamma ) $ plane the region that has  $\sigma ( gg \to \R \to \gamma \gamma ) \in (5,~10) $ fb. 
The vertical dashed line is the best fit width from the ATLAS.
For the dot-dashed and dashed horizontal lines, one has ${\rm Br} (\R \to \gamma \gamma ) =7 \times 10^{-4} $ and $1.0\times 10^{-3}$, which separately correspond to $\sigma(gg\to \gamma \gamma) =5~{\rm fb}$ and $\sigma(gg\to \gamma \gamma) =10~{\rm fb}$ for $\Gamma_{\rm tot}^\R =45$ GeV.  
We plot in the left panel of Fig.~\ref{widthi} contours of the $\Gamma(\R \to \gamma \gamma ) $ in the $g_X-\mathbf{v}_\R$ plane by assuming $M_X\approx 400~{\rm GeV}$, where the solid, dotted and dashed lines correspond to $\Gamma (\R \to \gamma \gamma ) =0.10~{\rm GeV}$, $0.05~{\rm GeV}$ and $0.02~{\rm GeV}$ respectively. 
It is clear that one needs ${\cal O } (1)$ quartic coupling $g_X$ and a large VEV $v_\R$ to enhance the decay rate. 
Large quartic couplings is helpful in keeping the vacuum stable up to the Planck scale, but might probably leads to a non-perturbative theory. 
One might need extra electroweak multiplets to enhance the diphoton rate for the case of small $g_X$. 
Apparently a wide width, $\Gamma_{\R}$, can not be explained if there is only $\gamma \gamma$ and $gg$ decay channels, which means there should be other decay channels of $\R$. 
Actually, if $2M_X< M_\R$, $\R$ might decay into di-scalars, whose decay rate can be written as
\begin{eqnarray}
\Gamma (\R \to 2 X) \approx  {g_X^2 \mathbf{v}_\R^2 \over 16\pi M_\R^2 } \sqrt{M^2 -4 M_X^2}  \; ,\hspace{1cm} X =\Theta, \Phi, \Omega
\end{eqnarray}
This rate can be very large for the light $X$, and be suppressed when $2 M_X  \approx M_\R$. 
We show in the right panel of Fig.~\ref{widthi}, contours  of the decay rate in the $M_X-g_X$ plane by setting $\mathbf{v}_\R =2.5~{\rm TeV}$.  
So the wide width of $\R$ can be explained without introducing new ingredients, but this scenario needs to be tuned and is thus less attractive. 
Alternatively If $\R$ can decay into the hidden valley, a wide $\Gamma_\R$ can naturally be explained as was studied in Ref.~\cite{Mambrini:2015wyu}.

Finally we  comment on the prediction of the scenario (ii). 
In this case one needs extra fermions to cancel the global  $SU(2)_L$ anomaly~\cite{globalsu2}, axial-vector anomaly~\cite{avector1,avector2,avector3} and the gravitational-gauge anomaly~\cite{anog1,anog2,anog3}.  
An economic set of new particles are three generations vector-like lepton doublets and singlets: $\psi_{L,R}^{\mathbf{1}}$ $(1,2, -1/2, \pm1 )$, $\psi_{L,R}^{\mathbf{2}}$ $(1,1, 0, \pm1 )$  and  $\psi_{L,R}^{\mathbf{3}}$ $(1,1, -1, \pm1 )$.  
New fermions might get mass through the Yukawa interactions with $\R$. 
The lightest neutral component of new fermions can naturally be the dark matter candidate, stabilized by the  local $U(1)_{B+L}$ gauge symmetry. 
As a result, there are extra contributions to the $\Gamma(\R\to \gamma \gamma)$ from new charged fermions, resulting in enhanced  $\Gamma (\R \to \gamma \gamma )$.
%
%
Furthermore $\R$ can naturally decay into dark sector in this scenario, which is helpful in deriving the best fit width.   
For the systematic study of symmetries behind the 750 GeV diphoton resonance, we refer the reader to Ref.~\cite{Chao:2015nsm} for detail.

\section{constraints and predictions}

\begin{figure}[t]
  \includegraphics[width=0.45\textwidth]{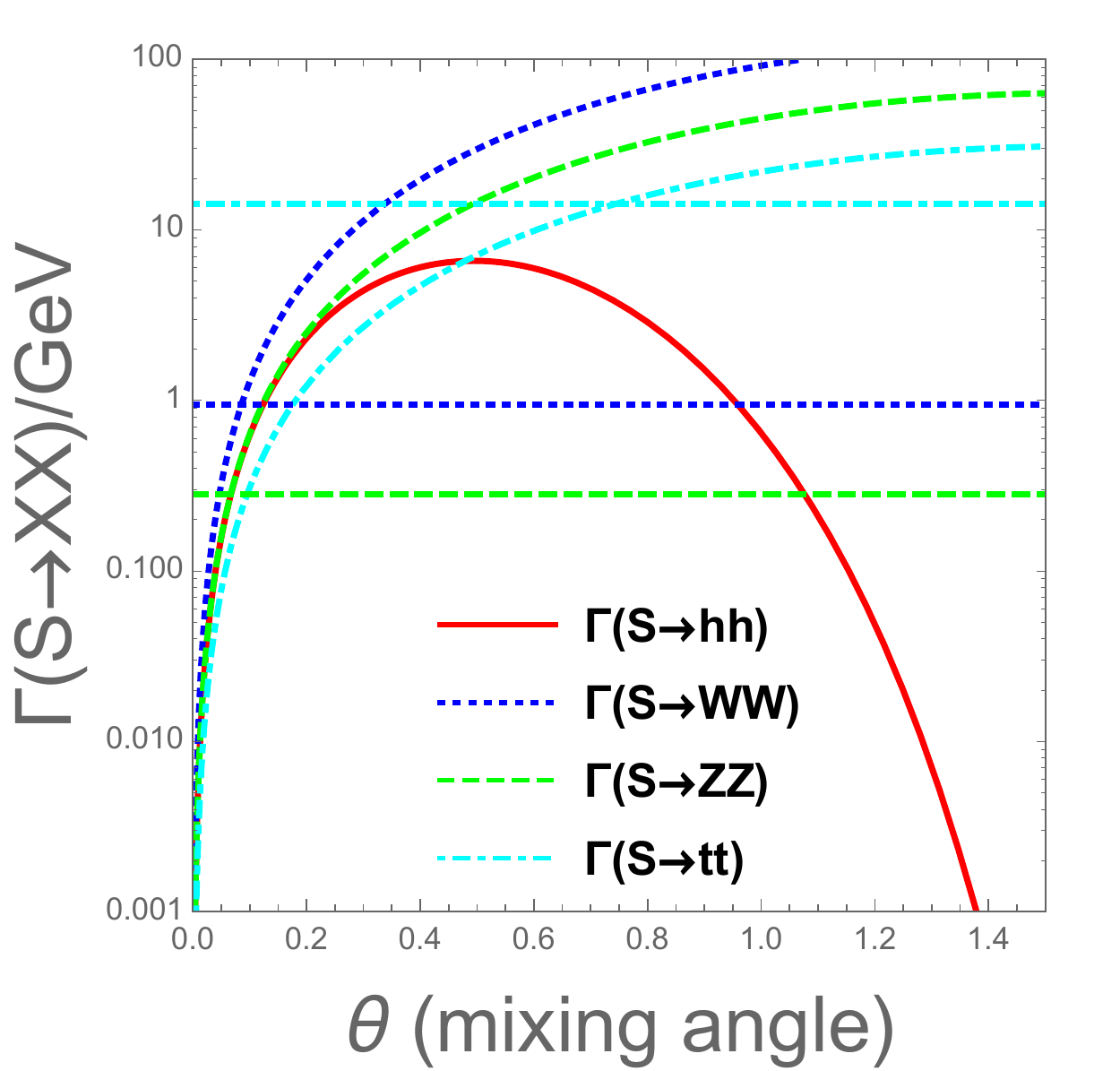}
   \includegraphics[width=0.45\textwidth]{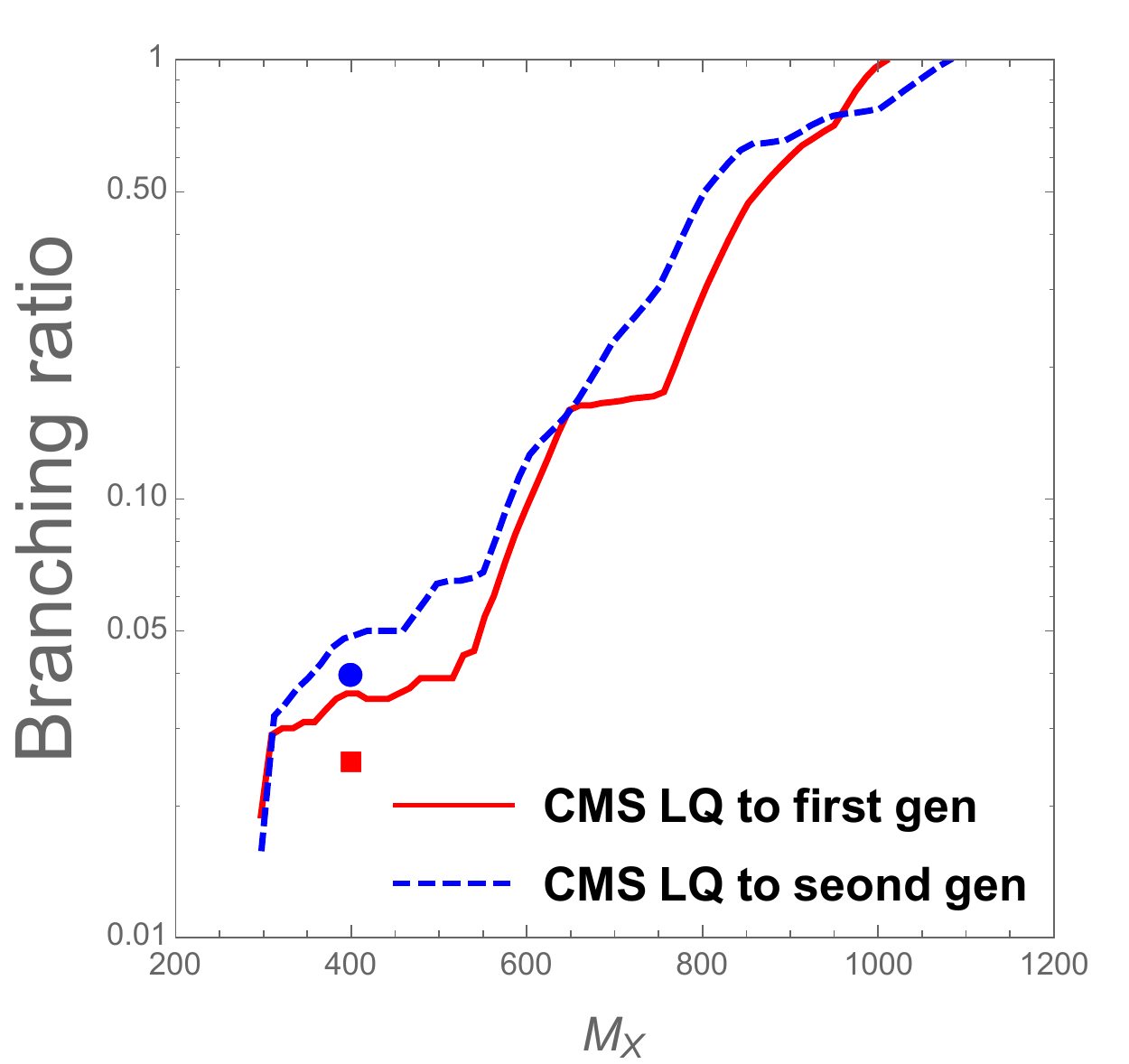}
\caption{\label{thetai}  Left panel:  constraint on the mixing angle between the new resonance and  the SM Higgs. The solid, dotted, dashed and dot-dashed lines correspond to $\Gamma(\R\to hh)$, $\Gamma(\R\to WW)$, $\Gamma(\R\to ZZ)$ and  $\Gamma(\R\to t\bar t)$ respectively; Right panel: Constraints on the leptoquarks from the CMS collaboration, the red square and the blue circle correspond to the case where  $\hat Y_X^{\rm 1gen}:\hat Y_X^{\rm 3gen}:\hat Y_X^{\rm 3gen}=1:1.25:6$.}
\end{figure}

The run-1 LHC has searched for the pair production of leptoquarks~\cite{Khachatryan:2015vaa}, which showed that the first and second generation scalar leptoquarks with the mass less than $1010 (850)$~${\rm GeV}$  are excluded for ${\rm BR} (X\to lq ) =1.0(0.5)$.  
For our case, leptoquarks couple to all three generation fermions, such that branching ratios of leptoquarks decaying  into the first and second generation fermions can be suppressed. 
We show in the right-panel of Fig.~\ref{thetai} the exclusion limit of leptoquark $X$  in the $M_X$- ${\rm BR}_X$ plane, given by the CMS~\cite{Khachatryan:2015vaa}, where the solid and dashed lines correspond to the exclusion limit of the  first and second generation leptoquarks respectively. 
By assuming the following ratio of Yukawa couplings $\hat Y_X^{\rm 1 gen}:\hat Y_X^{\rm 2 gen}:\hat Y_X^{\rm 3 gen}=1:1.25:6$, where $X=\Omega,\Phi$, we show in the right-panel of Fig.~\ref{thetai} the corresponding branching ratios of leptoquarks at $M_{X}=400~{\rm GeV}$. 
Apparently, they are consistent with the current collider bound of leptoquarks.
Alternatively leptoquarks may couple to the hidden valley resulting in  suppressed branching ratios of leptoquarks to SM fermions\footnote{We point out this possibility, but it does not have to be the realistic case.}. 
Embedding leptoquarks into supersymmetry where the mass eigenstate of  Bino ($\tilde B$) might be the dark matter candidate, one has the following new interactions: $ {\cal C}_1 \tilde B^T \tilde{\Omega}^c \Omega +  {\cal C}_2^{} \tilde B^T \tilde {\Phi}^c \Phi + {\rm h.c.}$
Then $X$ might decay into  $\tilde X$ and $\tilde B$.  
In this way, the current collider constraint will be invalid. 
A systematic investigation of collider signatures of leptoquarks at the LHC is  beyond the reach of  this paper and we leave it to a future study.

A second constraint on the model comes from diboson and dijet searches at the run-1 LHC with $\sqrt{s}=8~{\rm TeV}$, which have $\sigma\cdot {\rm BR }(ZZ) <12~{\rm fb}$~\cite{Aad:2015kna}, $\sigma\cdot {\rm BR }(WW) <40~{\rm fb}$~\cite{Aad:2015agg}, $\sigma\cdot {\rm BR }(hh) <39~{\rm fb}$~\cite{Aad:2015xja}, $\sigma\cdot {\rm BR }(t\bar t) <550~{\rm fb}$~\cite{ttbarx}. 
$\R$ might decay into these finial states through its mixing with the SM Higgs.  
By assuming $\Gamma(\R\to \gamma \gamma)\approx 0.05~{\rm GeV}$, one can derive the upper bound on the diboson and dijets decay rates, which on the other hand put upper bounds on the mixing angle $\theta$, describing the mixing between $\R$ and the SM Higgs.  
We plot in the left panel of  Fig.~\ref{thetai} various decay rates as the function of the mixing angle $\theta$, by setting $m_\R =750~{\rm GeV}$ and $v_\R=2.5~{\rm TeV}$. 
The solid, dotted, dashed and dot-dashed lines correspond to $\Gamma(\R\to hh)$, $\Gamma(\R\to WW)$, $\Gamma(\R\to ZZ)$ and  $\Gamma(\R\to t\bar t)$ respectively. 
Horizontal lines are corresponding upper bounds by assuming $\Gamma(\R \to \gamma \gamma ) =0.05~{\rm GeV}$.  
Apparently the $ZZ$ channel gives the strongest constraint, which has $\theta < 0.067$. 
It should be mentioned that this constraint depends on the value of $\Gamma(\R\to \gamma\gamma)_{\rm obs}$, a large $\Gamma(\R\to \gamma\gamma)_{\rm obs}$ will loose the constraint.

Finally we check gauge couplings unification in this model, which is an important motivation of extending the SM.
$\beta$-functions of gauge couplings can be written as 
\begin{eqnarray}
 \beta_{g_1} = {g_1^3 \over 16 \pi^2 } \left( {41\over 10 } + { 4 \over 3 } \right) \; , \hspace{0.5cm} \beta_{g_2} =-  {g_2^3 \over 16 \pi^2 }  {19 \over 6} \; , \hspace{0.5cm}
 \beta_{g_3} = -{g_3^3 \over 16\pi^2 } {19 \over 3 }  
\end{eqnarray} 
where the SM gauge couplings are normalized based on $SU(5)$ i.e., $g^\prime = \sqrt{3/5} g_1^{}$.  
Using the couplings, $\alpha_i \equiv g_i^2 /4\pi $, given as $(\alpha_1,~\alpha_2,~\alpha_3 ) =(0.01681,~0.03354,~0.1176)$ at the Z-pole, one might simulate the running behaviors of gauge  couplings, which shows that $g_1$ crosses  with $g_2$ at $\mu \approx 2.5 \times 10^{11}~{\rm GeV}$, and crosses with $g_3$ at $\mu \approx 6.5\times 10^{13}~{\rm GeV}$.
So there is no gauge couplings unification in this model, but one may approximately get the unification  immediately by extending the model with a quadruplet scalar with weak hyper charge $1/2$,  which might play important rules in generating Majorana neutrino masses via the modified type-III seesaw mechanism~\cite{Ren:2011mh}.  We leave the study of this part to a longer paper.

\section{conclusion}

A new resonance at invariant mass of $750$~GeV was observed by the run-2 LHC in the diphoton channel.
If confirmed it will be  a manifestation of new physics beyond the SM.
In this paper we investigated the possible explanation of this diphoton excess based on a TeV-scale neutrino mass model, that extends the standard model with four scalar singlets, two leptoquarks $\Omega$ and $\Phi$, one singly charged scalar $\Theta$ and one neutral scalar $\R$. 
Majorana neutrino masses are generated at the two-loop level. 
The diphoton excess is explained as the neutral scalar $\R$, which is produced at the LHC through the gluon fusion and decays into diphoton at the one-loop level with new charged  scalars running in the loop.
The value of $\Gamma (\R\to gg)/\Gamma(\R \to \gamma \gamma )$ is about ${\cal O } (25)$, which is the typical character of this model. 
It is greatly enhanced comparing with the conventional vector-like quark models.   
Interestingly the model fits perfectly with a wide width of $\R$.  
Constraints on the model was studied, which show a negligible mixing between $\R$ and the SM Higgs.  
Besides, leptoquarks need to couple with the hidden valley so as to avoid constraints from the LHC run-1. 
There is no gauge couplings unification (GU) in this model, but  GU can be approximately derived by extending the model with a scalar quadruplet. 
We expect the future high luminosity  run-2 LHC could shed light on  the resonance in other channels, and show us the gate of new physics.

\begin{acknowledgments}
The author thanks to Huai-ke Guo, Ran Huo, Hao-lin Li, Grigory Ovanesyan  and Jiang-hao Yu for very helpful discussions. 
This work was supported in part by DOE Grant DE-SC0011095.
\end{acknowledgments}

\end{document}